\documentclass[journal]{IEEEtran} %,draft
\ifCLASSINFOpdf \else \fi

\usepackage{graphicx}

\usepackage[cmex10]{amsmath}
\usepackage{amsbsy}
\usepackage{amssymb}
\usepackage{amsthm}
\usepackage{setspace}

\usepackage{mathtools,lipsum, nccmath}

\usepackage{cuted}
\usepackage{algorithmic}
\usepackage{array}
\ifCLASSOPTIONcompsoc
 \usepackage[caption=false,font=normalsize,labelfont=sf,textfont=sf]{subfig}
\else
 \usepackage[caption=false,font=footnotesize]{subfig}
\fi
\usepackage{url}
\usepackage{hyperref}
\RequirePackage{threeparttable}
\RequirePackage{booktabs}
\RequirePackage{multirow}
\usepackage{tabularx}
\usepackage[noadjust]{cite}
\usepackage[numbers,sort&compress]{natbib}

\usepackage[none]{hyphenat}

\usepackage{titlesec}

\titleformat{\paragraph}
{\normalfont\normalsize\bfseries}{\theparagraph}{1em}{}
\titlespacing*{\paragraph}
{0pt}{3.25ex plus 1ex minus .2ex}{1.5ex plus .2ex}

\theoremstyle{definition}

\title{Estimation of Building Energy Demand Characteristics using Bayesian Statistics and Energy Signature Models}

\date{March 2025}

\author{
   \IEEEauthorblockN{Justinas Smertinas\IEEEauthorrefmark{1}, 
                    Nicolaj Hans Nielsen\IEEEauthorrefmark{1},
                    Matthias Y. C. Van Hove\IEEEauthorrefmark{1},
                    Peder Bacher\IEEEauthorrefmark{1}, 
                    Henrik Madsen\IEEEauthorrefmark{1}}\\
    \IEEEauthorblockA{\IEEEauthorrefmark{1}Department of Applied Mathematics and Computer Science, Technical University of Denmark, Denmark
    \\\{jussm, s184335, mvaho, pbac, hmad\}@dtu.dk}\\
\thanks{\textit{Corresponding author at:} Asmussens Allé, 303B, 019, 2800 Kgs. Lyngby Denmark}% <-this % stops a space
\thanks{\textit{E-mail:} \href{mailto:jussm@dtu.dk}{jussm@dtu.dk} (Justinas Smertinas)}}

\begin{document}
\maketitle
\markboth{Draft}%
{}

%------- Including all the sections ----------
%---------------------------------------------
\begin{abstract}
\noindent This work presents a scalable Bayesian modeling framework for evaluating building energy performance using smart-meter data from 2,788 Danish single-family homes. The framework leverages Bayesian statistical inference integrated with Energy Signature (ES) models to characterize thermal performance in buildings. This approach quantifies key parameters such as the Heat Loss Coefficient (HLC), solar gain, and wind infiltration, while providing full posterior distributions to reflect parameter uncertainty.

Three model variants are developed: a baseline ES model, an auto-regressive model (ARX-ES) to account for thermal inertia, and an auto-regressive moving average model (ARMAX-ES) that approximates stochastic gray-box dynamics. Results show that model complexity improves one-step-ahead predictive performance, with the ARMAX-ES model achieving a median Bayesian R² of 0.94 across the building stock. At the single-building level, the Bayesian approach yields credible intervals for yearly energy demand within $\pm1\%$, enabling more robust diagnostics than deterministic methods.

Beyond improved accuracy, the Bayesian framework enhances decision-making by explicitly representing uncertainty in building performance parameters. This provides a more realistic foundation for investment prioritization, demand forecasting, and long-term energy planning. The method is readily applicable to other building typologies or geographies, offering a scalable tool for data-driven energy management under uncertainty.
\end{abstract}

\section{Introduction}
\noindent Buildings are one of the largest sources of energy use in Europe, accounting for more than 40\% of all energy consumption in Denmark and approximately 39\% worldwide \citep{Energistyrelsen_2018_40percent, SOMU2020114131}. Boosting building energy efficiency will have numerous downwind consequences. It would cut emissions, tackle energy poverty, reduce people's vulnerability to energy prices, and benefit the health and well-being of inhabitants, among others. In order to achieve a fully decarbonized building stock by 2050, the Energy Performance of Buildings Directive (EPBD) was introduced in 2002 \citep{EPBD2002} and contributes directly to the long-term goals of the EU with energy and climate.

Since 2012, the new directive on energy efficiency \citep{EED2012} states that \textit{'all member states are responsible for the installation of individual energy meters, including heat meters, in all buildings to the extent that it is technically possible and economically feasible'}. The growing implementation of energy monitoring and smart metering systems has resulted in an increased availability of long-term measurements of real energy use in occupied buildings. In particular in Scandinavia, the penetration of the smart meters market is approaching 100\% \citep{JRC_meter_rollout}, generating extensive datasets for, for example, energy reduction strategies.

Today, a large number of studies in EU countries have been published, focusing on the difference between the calculated regulatory energy use and the real energy use \citep{Sunikka_2012, Majcen_2013, Delghust_2015, Dronkelaar_2016, MVH_EPC, MVH_phd, DEWILDE201440}. All report that the reliability of the regulatory calculation methods and their energy labels is limited. In order to improve energy labeling, a number of those studies point towards (simple) data-driven methods for thermal energy performance estimation \citep{Delghust_2015, Dronkelaar_2016, MVH_EPC, MVH_phd}. Today, with the availability of high-frequency energy monitoring data, the search for scalable data-driven methods for building stock evaluation has become more relevant than ever. The present paper presents exactly such a data-driven and scalable approach.

\subsection{Literature review}

\noindent The current state-of-the-art statistical approaches for building energy demand forecasts include a wide range of frequentist methods, such as Artificial Neural Networks (ANNs), Support Vector Machines (SVM), time-series approaches such as ARIMAX, or stochastic-differential-equation-based Grey-Box models, and finally deep learning approaches and ensemble methods\citep{SUN2020110022, CHEN20222656, GRILLONE2020110027}. These approaches are often used due to their scalability and adaptability to high-frequency data streams, which is an invaluable asset in energy modeling.
However, due to the shared Frequentist methodology, these methods (by definition) assume that the parameters of a model are fixed but unknown quantities \citep{gelman1995bayesian}. This can lead to overconfidence in model predictions and parameter estimates, especially in complex real-world scenarios where the data is noisy and incomplete, leading to sub-optimal decisions \citep{kahneman1982judgment_overconfidence}.

The Frequentist model strengths lie in their ability to be trained on large datasets and optimized for predictive accuracy, which is invaluable for energy modeling, however, they are less suited for direct uncertainty quantification, another essential aspect of energy modeling in practice. The challenge with statistical studies lies partially in choosing the correct statistical methodology for the task at hand. When the task involves building an evaluation from inferred model parameters, an alternative methodology may be more suited to the task.\\

\subsubsection{A Case for Bayesian Methods in Building Energy Modeling}\label{sec:Backg_Bay_Insights}

\noindent In contrast, Bayesian statistics offers an alternative approach to the widespread Frequentist approach in building energy modeling \citep{GRILLONE2020110027}. Bayesian statistics focuses on the probability distribution of inferred model parameters rather than the model's predictive performance.

The main feature of the Bayesian approach is that it is suited for creation of robust and informative models, especially in contexts where data is noisy or assumptions about the underlying distribution may not hold. This robustness comes at a computational cost limiting complexity of model which can be considered, however, as this paper demonstrates, the trade-off is easily justifiable. In Bayesian modeling, rather than focusing on a single parameter value that maximizes predictive performance, the approach estimates the posterior distribution of the parameters, which provides a more complete picture of their uncertainty.\\

\noindent For the sake of brevity, the explicit difference between the Frequentist and Bayesian approaches is omitted, and instead the focus in the following section is on presenting a holistic understanding of the trade-off between the two approaches. \\

\subsubsection*{Bayesian Inference: Prior and Posterior Distributions}

\noindent The goal of the Bayesian framework is to update prior beliefs about model parameters with observed data \citep{B_Data_Analysis_2}. In case of this paper, updating prior beliefs about building characteristics after observing the building's smart-meter data. This is done through Bayes' theorem:

\begin{equation}
    p(\theta|X) = \frac{p(X| \theta)p(\theta)}{p(X)}
\end{equation}
where:
\begin{itemize}
    \item $p(\theta)$ - \textit{the prior}, which encapsulates the prior knowledge or assumptions about the parameters before observing the data.
    \item $p(\theta|X)$ - \textit{the posterior distribution}, which represents the updated beliefs about the model parameters $\theta$ after observing the data $X$.
    \item $p(X|\theta)$ - \textit{the likelihood}, representing the probability of the observed data given the parameters. Specified with the help of the statistical model of choice.
    \item $p(X)$ - \textit{the marginal likelihood or evidence}, which normalizes the posterior but is often not of primary concern for model estimation.
\end{itemize}

\noindent Bayesian model fitting is performed by sampling the posterior distribution. Each component of the posterior parameter distribution plays a role in this process. \\

\subsubsection*{Incorporating Prior Knowledge}

\noindent For instance, the \textit{prior distribution} $p(\theta)$ allows to incorporate prior expert knowledge about model paraneters, e.g. Heat Loss Coefficients (HLCs) for certain types of buildings. This knowledge can be encoded in the model to reflect reasonable expectation of reality (e.g. that HLCs for Danish single-family homes are around (some number) and can be described by a Gamma distribution, since negative HLCs are not physically possible.). \textit{This ensures that the model's estimates are physically interpretable and informed by reality, even before any new data is introduced.} As more data becomes available, the influence of the prior decreases, and the posterior distribution $p(\theta | X)$ provides an update with more accurate understanding of the parameters, including any uncertainties around them.\\

\subsubsection*{Directly Quantifying Uncertainty}

\noindent One of the most valuable aspects of the Bayesian methods is the ability to directly quantify uncertainty through the posterior distribution. For example, rather than offering a point estimate for a parameter like HLC, Bayesian methods give a full probability distribution. This distribution \textit{shows the range of likely values for the parameter, which allows better decision making.} Credibility intervals, derived from the posterior distribution, may give a more informative measure of uncertainty than traditional confidence intervals. This is crucial in real-world building energy modeling, where the assumptions necessarily made in frequentist models often lead to over-confident estimates.

Frequentist methods can produce confidence intervals that shrink asymptotically as more data is added, even if the assumptions about normality or linearity are violated. In contrast, Bayesian credibility intervals reflect the true uncertainty of the parameters and do not shrink artificially when the underlying assumptions of the model are incorrect. Most notably, the Bayesian Credibility Interval is a statement of probability about the parameter value given fixed bounds, while the Frequentist Confidence Interval is a probability about the bounds given a fixed parameter value \citep{HESPANHOL2019290}. This ability to more directly quantify uncertainty leads to more informative building characteristic estimation, more reliable forecasts, and enables better energy efficiency planning, while avoiding the pitfalls of overconfidence.

\subsection{Applying Bayesian approach on Energy Signature}

\noindent The Bayesian approach is simply an alternative way of describing the statistical model, it assumes an underlying statistical model. This present work suggests a Bayesian framework to the well-established Energy Signature models. The reasons for the persistence of the ES model in building engineering are numerous. Even in its most basic form, the ES model offers a clear physical interpretation, which allows it to accurately represent the heat balance equation \citep{rasmussen2020method}. One of the key strengths of the ES model is its expandability. For example, it can be enhanced with B-splines to estimate the azimuth angle-dependent solar gain factor, thus incorporating the effects of solar radiation on heating demand \citep{RASMUSSEN_solar}. Furthermore, periodic B-splines can be introduced to capture daily demand periodicities, providing a more refined model of the building energy use cycle \citep{Justinas-Splines}. Additionally, the ES model can be extended with auto-regressive (AR) and moving-average (MA) with eXogenous input components, transforming it into an ARMAX form that acts as a discrete-time approximation of physical state-space models. This link bridges the gap between data-driven methods and gray-box physics-based models, making the ES model a robust choice to model complex building dynamics \citep{ARMAX_physical}.

\subsection{Purpose of this research}

\noindent The purpose of this research is to advance the methodology for evaluating building energy performance through the integration of Bayesian statistical modeling within the well-established Energy Signature (ES) framework. Using Bayesian methods, this study aims to address potential oversights in traditional energy performance assessments, including overconfidence in parameter estimates and the inability to robustly quantify uncertainty.

The research focuses on practical implications, with two main objectives:

\begin{enumerate}
    \item \textbf{Demonstrate a scalable probabilistic energy modeling framework:} The present work demonstrates the versatility of the Bayesian ES model in accommodating both individual building assessments and portfolio-scale evaluations. By providing probabilistic estimates of key parameters, such as Heat Loss Coefficients (HLC), the framework supports reliable and interpretable insights, from single buildings to large housing stocks.
    \item \textbf{Complement traditional energy certification frameworks:} By demonstrating the utility of a Bayesian approach to refine and supplement existing energy performance certification methods, this work offers a scalable tool for evaluating energy performance that adapts to the availability of data from high-frequency smart meters. This capability bridges the gap between regulatory requirements and data-driven innovations.
\end{enumerate}

\noindent Finally, this work seeks to introduce large-scale probabilistic (Bayesian) building energy modeling by evaluating n=2788 Danish single-family homes. The novel approach offers valuable findings and actionable insights for building owners, engineers, policymakers and energy-providers alike.

%\noindent This research is intended for practical adoption. The flexibility of the Bayesian framework enables applications on both individual and portfolio scales, allowing engineers to accurately evaluate single buildings and entire housing stocks. For building owners, probabilistic estimates support realistic cost evaluations, empowering investment in targeted energy-saving upgrades. For engineers, this tool not only updates traditional EPC assumptions, but also provides a scalable, data-driven alternative for creating building digital twins and identifying low hanging fruits in energy efficiency - revealing buildings that could yield high returns on modest energy-saving investments.

%Lastly, this work demonstrated the feasibility of the Bayesian ES model as a modern complement to deterministic methods in building certification. By offering a dynamic, evidence-based approach, this framework has the potential to improve both individual building assessments and broader strategies to achieve decarbonized and energy-efficient building stock.

\subsection{Structure of this Paper}

\noindent This paper is organized as follows: Section \ref{sec:Background} provides an overview of the Energy Signature (ES) model, including its historical context, fundamental principles, and contemporary adaptations for improved energy performance assessments. Section \ref{sec:Method} outlines the data sources and methodology, describing the Bayesian modeling framework and its integration with ES models. Section \ref{sec:Results} presents the results, including model validation, performance metrics, and a detailed analysis of posterior parameter distributions. A single-building case study is also included to illustrate the practical implications of Bayesian inference for energy diagnostics. Section \ref{sec:Discussion} discusses the broader implications of the findings for building owners, engineers, and policymakers, as well as limitations and future research directions. Finally, Section \ref{sec:Conclussion} concludes the paper by summarizing key insights and emphasizing the potential of Bayesian methods in building energy performance assessments.

\section{Background}\label{sec:Background}

\subsection{Introduction to Energy Signature (ES) Model}

\noindent To demonstrate the added informational value of the Bayesian approach in the field of digital building diagnostics, it is applied on the well-tested Energy Signature (ES) model for building's heat demand. ES models are widely applied data-driven models that express heating energy use as a function of weather variables. They have been applied at least since the early 1950s \citep{first_ES} and are still actively developed and applied today \citep{rasmussen2020method, real2021revealing, RASMUSSEN_solar}. To understand the full range of models considered in this work, the foundational concepts of the contemporary ES model are introduced in this section.\\

\subsubsection*{Energy Signature Model Basics}

\noindent The dominating principle of the method is to apply linear regression, to explain a simple energy-balance equation, like

\begin{equation}\label{eq:Heat_Balance}
    \Phi_{\text{heat}} - \Phi_{\text{tr}} + \Phi_{\text{sol}} + \Phi_{\text{int}} - \Phi_{\text{vent}} +\Phi_{\text{mass}} + \Phi_{\text{latent}} = 0 
\end{equation}

\begin{figure}[h]
    \centering
    \includegraphics[width=0.85\columnwidth]{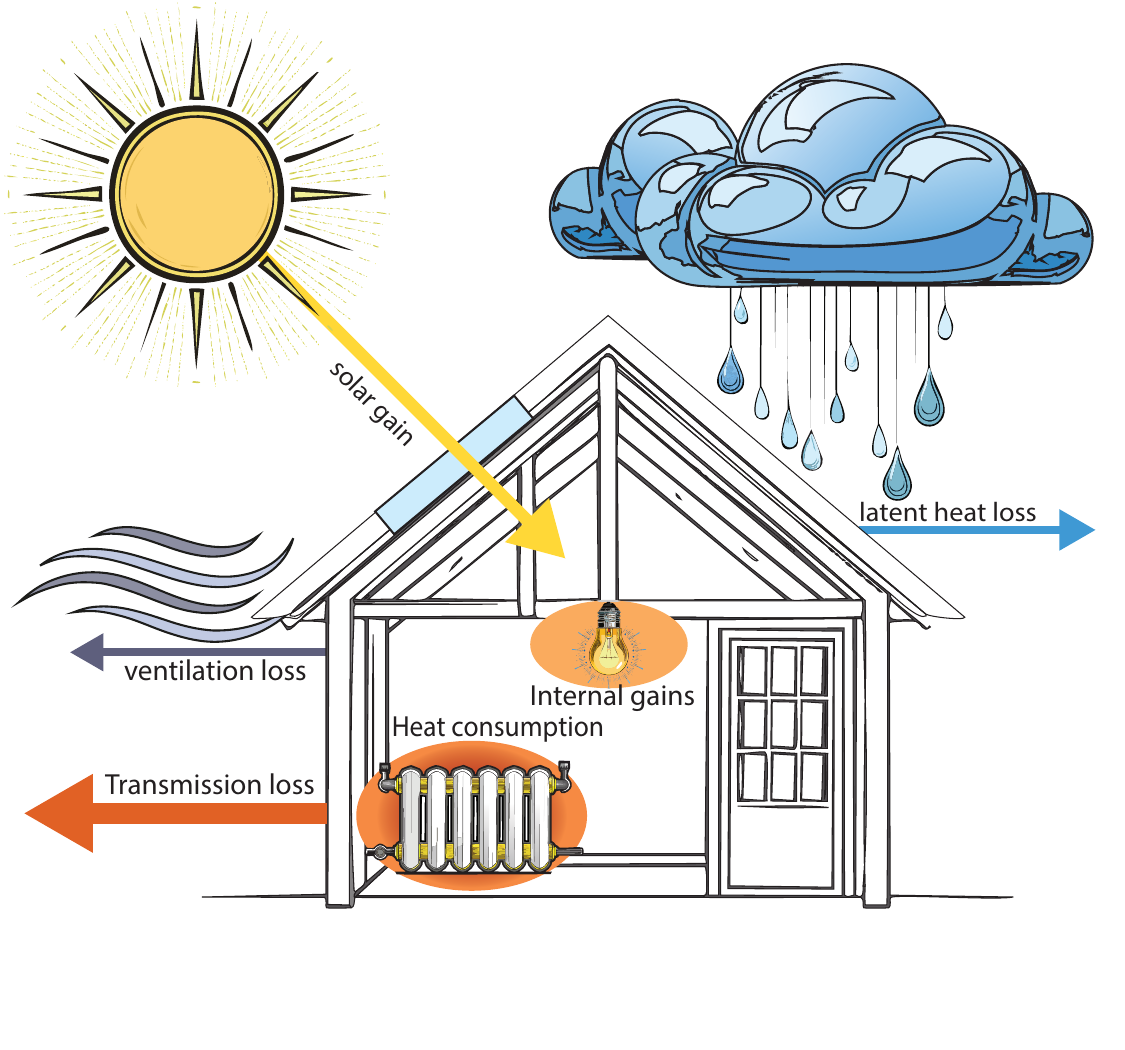}
    \caption{Visualisation of a building's heat balance.}
    \label{fig:Intro_Heat_Equation}
\end{figure}

\noindent where $\Phi_{\text{heat}}$ is the heat consumption, $\Phi_{\text{tr}}$ is the transmission loss, $\Phi_{\text{sol}}$ is the solar gain, $\Phi_{\text{int}}$ is the internal heat gains, $\Phi_{\text{vent}}$ is the ventilation loss, $\Phi_{\text{mass}}$ is the release of thermal energy stored in the thermal mass and $\Phi_{\text{latent}}$ is the absorption and release of energy due to evaporation and condensation in the thermal zone \citep{rasmussen2020method}. Equation \ref{eq:Heat_Balance} is visualized in Figure \ref{fig:Intro_Heat_Equation}.

The original ES model is constructed by lumping $\Phi_{\text{sol}}$, $\Phi_{\text{int}}$ and $\Phi_{\text{vent}}$ into a single parameter $\Phi_{0}$ and using the simplification that $\Phi_{\text{tr}}$ consists only of transmission loss to the outdoor air. With this simplification, the ES model becomes

\begin{equation}
   \overbrace{\Phi_{\text{heat}} = \Phi_{0} + \underbrace{ \text{UA}_0 \cdot (\mathbf{T}_{\text{indoor}}-\mathbf{T}_{\text{outdoor}})}_{\Phi_{\text{tr}} \text{ simplified}} + \epsilon_{t}}^{\text{Original Energy Signature Model}}
\end{equation}

\noindent with $\Phi_{\text{heat}}$ the energy use for space heating [W],  $(\textbf{T}_{\text{indoor}}-\textbf{T}_{\text{outdoor}}$) [K] is the difference between indoor and outdoor temperatures, $\text{UA}_0$ the relation of the energy use for space heating with the outdoor temperature [W / K]. \citep{rasmussen2020method}\\

\noindent The equation is commonly visualized as a linear function, with a slope and an intercept, in a 2-dimensional graph. The slope, $\text{UA}_0$, represents the relationship of energy use for space heating and is often also referred to as the building Heat Loss Coefficient (HLC). The intercept $\Phi_{0}$, refers to the weather-independent part of energy use for space heating and is modeled as a constant for buildings without cooling and heat recovery. Note that this model is applied only on observations from the part of the year when the building is actively heated. \\

\subsubsection*{The Contemporary Energy Signature}\label{sec:Compemporary_ES}

\noindent The original ES model formulation has a few clear drawbacks. One, it only considers building observations during the heated part of the year, and second, is that buildings' energy demand is assumed to only be dependent on the relationship with the outdoor temperature. The first drawback is a hindrance for a direct way to estimate yearly heat consumption with the ES model. The second is that by not accounting for other significant weather influences this introduces bias into the HLC estimate. Thus, to account for these, the ES models have been actively developed since the 1950s, with several key improvements.\\

\noindent The first improvement extends the original ES model to being able to describe energy consumption throughout the entire year, and not only during explicitly heated periods.

To achieve this, the contemporary ES versions \citep{rasmussen2020method} express the heat demand equation as a smooth mixture between two regimes governing heat demand during the changing seasons, namely a \textit{Winter Regime} ($\Phi_{\text{winter}}$) and \textit{Base-Heat-Load Regime} ($\Phi_{\text{base}}$). The model transitions smoothly between the two regimes via the Log-Sum-Exponential (LSE) function, which is a type of smooth-max function.

\begin{equation}
    \Phi = \text{LSE}_k(\Phi_{\text{winter}}, \Phi_{\text{base}})
\end{equation}

\noindent The benefit of using the LSE function for switching between the two regimes is that it allows to tailor the smooth regime switching sensitivity for each building individually at a cost of estimating only one additional parameter ($k$).

It is important to note that this is not the only way to achieve a mixture of regimes spanning two seasons. It is simply the method which testing shows works best in the case of Danish district heating, where the heat exchangers are usually turned on automatically. For more abrupt transitions, a Gaussian mixture model should be considered \citep{nielsen2024a}. \\

\begin{figure*}[!h]
    \centering
    \includegraphics[width=\textwidth]{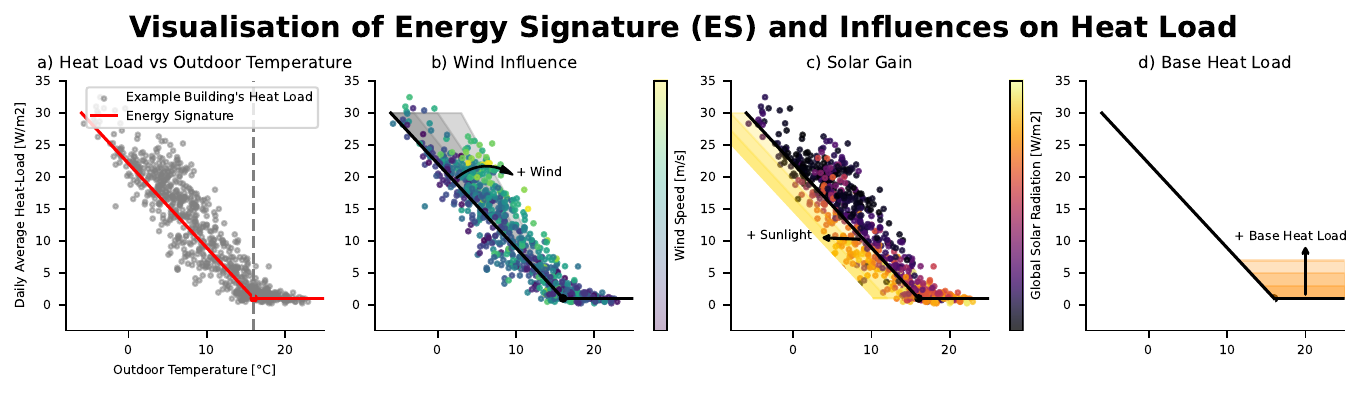}
    \caption{a) Visualisation of the Energy Signature model overlayed on daily-average heat-load observations for an example building. b) Visualisation of the wind's influence on heat loss. c) Visualisation of solar gain. d) Visualisation of increased base heat-load.}
    \label{fig:ES_conceptualisation}
\end{figure*}

\noindent As mentioned above, the energy demand of buildings is influenced by a variety of external factors.  Figure \ref{fig:ES_conceptualisation} shows a conceptual illustration of external forces driving the change in heat consumption in the ES model. For example, the wind speed increases the heat loss caused by the increased air exchange as shown in \citep{Christoffer_airtightness}. Ultimately, this leads to a steeper gradient of the energy signature function as indicated in Subfigure \ref{fig:ES_conceptualisation}b (Wind). It is easily understood that solar gain alters the base temperature of the building and heating up internal air temperature while maintaining the HLC of the building, which is a physical property. This is illustrated in Subfigure \ref{fig:ES_conceptualisation}c (Solar).

In general, it is also not feasible to obtain an indoor temperature measurement for every building equipped with a smart meter. This issue can be circumvented by replacing $\mathbf{T}_{indoor}$ in the original ES model with an estimated parameter for the base temperature ($\text{T}_{base}$) in which the building is thermally in equilibrium with the ambient air. This parameter can also be interpreted as the temperature at which the building's heating system switches on the heating.

In explicitly mathematical terms, to include all the aforementioned improvements, the model regimes and influence of external weather factors for the contemporary ES model are parametrized as follows:

\begin{equation}\label{eq:ES_1}
    \begin{split}
    \mathbf{\Phi}_{\text{winter}}^{(t)} & = \underbrace{(\text{UA}_0 + \text{UA}_{\text{wind}} \mathbf{W}_{\text{speed}}^{(t)})}_\text{Effective Heat-Loss Coefficient (HLC)} (\text{T}_{\text{base}} - \mathbf{T}_{\text{outdoor}}^{(t)})\\
    & - \underbrace{\text{gA } \mathbf{I}_g^{(t)}}_\text{solar gain}\\ 
    & + \Phi_0 + \epsilon^{(t)}\\
    \mathbf{\Phi}_{\text{base}}^{(t)} & = \Phi_0 + \epsilon^{(t)}
    \end{split}
\end{equation}

\noindent In equation \ref{eq:ES_1} the explanatory data variables are: $\textbf{W}_{\text{speed}}^{(t)}$ - the wind speed, $\textbf{I}_g^{(t)}$ is the global solar irradiation, and $\textbf{T}_{\text{outdoor}}^{(t)}$ is the ambient temperature at time $t$. The model parameters are $\text{UA}_0$ - heat loss coefficient under wind-free conditions \citep{HLC_def}, $\text{UA}_{\text{wind}}$ is the wind dependency, $\text{T}_{\text{base}}$ is the base temperature at which the building is in thermal equilibrium, $\text{gA}$ is the solar transmittance, and $\phi_0$ is the standby effect or rather base heat load. Lastly, $\epsilon^{(t)}$ and $\xi^{(t)}$ are the error terms.\\

\subsection{Bayesian Model Fitting and Evaluation}

\noindent The primary approach for statistical data analysis in this paper is the Bayesian approach. As this is not a typical methodology in the field of energy performance of buildings, the following subsection presents the key concepts required to understand the methodology and result evaluation, such as the necessary model evaluation metrics.

In Bayesian statistics, a probability model is fitted to a dataset ($\textbf{X}$), with the goal of estimating a probability distribution on the model parameters ($\theta$) and on other values, such as predictions for unobserved data \citep{gelman1995bayesian}. The desired posterior distributions for model parameters are usually obtained through an involved Markov Chain Monte Carlo (MCMC) sampling procedure.

The posterior predictive distribution, denoted $p(\Tilde{\textbf{X}}) = p( \Tilde{\textbf{X}} | \textbf{X})$, plays a crucial role in the validation of the Bayesian model. This process, commonly referred to as posterior predictive checking, is fundamental to the Bayesian approach, as performing inference on a deficient model would be meaningless. 

The procedure begins by sampling $M$ instances from the inferred posterior parameter distribution, denoted $\theta^{(m)} \sim p(\theta | \textbf{X})$. Subsequently, for each parameter $\theta^{(m)}$ a sample from the posterior predictive is drawn as $\Tilde{\textbf{X}}^{(m)} \sim p(\textbf{X}|\theta^{(m)})$. In the context of this paper, this can be thought of as a single simulated heat-demand scenario given $\theta^{(m)}$ -- the values of the sampled model parameters.

In an ideal scenario, the model should generate samples of simulated heat demand $\Tilde{\textbf{X}}^{(m)}$ that are indistinguishable from the observed heat-demand data $\textbf{X}$ \citep{pml1Book}. However, achieving this ideal outcome is often impractical. Therefore, the guiding principle is to ensure that the model captures the most important data characteristics for the specific application \citep{lambert2018student}. This section introduces a few metrics by which the models in this paper will be evaluated.\\

\subsubsection*{Expected Log Pointwise Predictive Posterior (ELPD) and Leave One Out Cross-Validation}

\noindent In Bayesian model fitting, the objective is to minimize the negative log-posterior. Therefore, the natural and default in the field metric of concern would be the expected log pointwise predictive distribution (ELPD),

\begin{equation}
    \text{ELPD} = \Sigma_{i=1}^n \int p_t (\Tilde{x}_i)\log p(\Tilde{x}_i|\textbf{X})d \Tilde{x}_i
\end{equation}

\noindent where $p_t(\Tilde{x}_i)$ is the density of an unseen dataset, $\textbf{X}_{\text{new}}=(\Tilde{x}_1, ..., \Tilde{x}_n)$. However, the distribution of unseen data, $p_t(\Tilde{x}_i)$ is, of course, unknown. Therefore, only an approximate quantity for log-posterior density can be computed,

\begin{equation}
    \text{LPD} \approx \sum_{i=1}^{n} \log (\frac{1}{M}\sum_{m=1}^M p(x_i| \theta^{m}))
\end{equation}

However, since this will be an optimistic approximation of the ELPD, instead, a leave one out cross-validation is employed as demonstrated by \citep{nielsen2024a}. 

Let $\textbf{X}_{-i}$ denote the vector of all except the $i^{th}$ observation, and conversely $x_i$ be the $i^{th}$ observation. Then the leave one out ELPD is

\begin{equation}
    \text{ELPD}_{\text{loo}} = \sum_{i=1}^n \log p(x_i\mid \textbf{X}_{-i})
\end{equation}

where 

\begin{equation}
    p(x_i \mid \textbf{X}_{-i}) = \int p(\textbf{X}_i\mid \theta) p(\theta \mid \textbf{X}_{-i}) d\theta
\end{equation}

is the leave-one-out predictive density given the data without the $i^{th}$ observation \citep{nielsen2024a}. But, since evaluating this expression would require retraining the model $N$ times, this is exceptionally costly in the Bayesian framework. Instead, an approximate quantity \citep{37_gelfand} is used, which is implemented in most Bayesian modeling frameworks such as \texttt{PyMC} \citep{abril2023pymc}. \\

\subsubsection*{Bayesian R-squared}

\noindent The classical $R^2$ statistic is a well-known and widely used measure of model fit in linear regression, representing the proportion of variance in the outcome explained by the model. However, its direct application in the Bayesian context is problematic. First, when using posterior samples instead of point estimates, the variance of fitted values can exceed the total variance in the data, leading to $R^2 > 1$, which is conceptually incoherent. Second, the classical formulation ignores the uncertainty inherent in Bayesian inference.

To address these issues, Gelman et al.~\citep{Bayes_R2} propose a Bayesian version of $R^2$, grounded in a variance decomposition that reflects both model fit and residual uncertainty. The Bayesian $R^2$ is defined as:
\begin{equation}
R^2_{\text{Bayes}} = \frac{\text{Var}(\mathbb{E}[y \mid \theta])}{\text{Var}(\mathbb{E}[y \mid \theta]) + \mathbb{E}[\text{Var}(y \mid \theta)]}
\end{equation}
where both numerator and denominator are computed for each posterior draw $\theta$, and then summarized over the posterior. The numerator represents the variance of the model’s predicted means (explained variance), and the denominator includes both the explained variance and the expected residual variance, ensuring that $R^2_{\text{Bayes}} \in [0, 1]$ by construction.

This formulation ensures that Bayesian $R^2$ is compatible with the principles of Bayesian inference and remains interpretable as the proportion of variance in new data explained by the model. Unlike its frequentist counterpart, this version incorporates uncertainty and can be visualized as a posterior distribution, allowing for richer diagnostic insight into model performance.\\

\subsubsection*{Bayesian \textit{p}-value and testing parameter significance}

\noindent The Bayesian $p$-value is similar to that in the frequentist approach. In both approaches, in order to ensure that the model and the posterior parameter have certain characteristics, a test is introduced. The idea is first to introduce a test statistics that encompasses the important characteristics, $T(\cdot)$, and then compare the evaluation on the data, $T(\textbf{X})$, with the evaluation on the generated samples, $T(\Tilde{\textbf{X}})$. The test statistic can also be used to formulate a flexible class of p-values, including tests to prove practical significance \citep{nielsen2024a}. With the test statistic, the Bayesian p-values can be formulated as

\begin{equation}
    \begin{split}
        p_T = P(T(\Tilde{\textbf{X}}, \theta) \geq T(\textbf{X}, \theta) | \textbf{X})\\
    \end{split}
\end{equation}

In \citep{mackay2003information}, it is demonstrated that there are a variety of ways to construct such useful tests, in particular that would aid in practical domain knowledge check. 

However, the direct approach to assess parameter significance is to calculate the $95\%$ credibility interval based on $p(\theta | \textbf{X})$ and check whether the interval includes 0. If so, the corresponding parameter is considered insignificant. \\

\section{Materials and methods}\label{sec:Method}

\subsection{Data}

\subsubsection{Building Data overview}

\noindent The dataset utilized for this study originates from Aalborg University (AAU) and consists of smart heat meter data collected from residential buildings within Aalborg Municipality, Denmark. The data spans from 2018 to 2022 and includes detailed building characteristics sourced from the Danish Building and Dwelling Register (BBR) and Energy Performance Certificate (EPC) input data \citep{AAU_data}.

The dataset comprises measurements from 34,884 commercial smart heat meters and 10,765 commercial smart water meters, providing high-resolution hourly data on heat and water consumption. The smart meter data has been systematically processed to ensure consistency, removing erroneous or missing values. Additionally, the EPC data includes up to 86 distinct characteristics per building, offering comprehensive insights into building performance and thermal efficiency

For the purpose of this study, we focus exclusively on the EPC-labeled subset of buildings (n=2788). This selection will enable future evaluation of building energy performance estimated from Bayes-ES within the framework of energy certification, allowing for direct comparisons between estimated and measured heat demand. The dataset serves as a foundation for the Bayesian modeling approach applied in this work, supporting the quantification of uncertainties in building energy assessments.

\begin{figure}[h]
    \centering
    \includegraphics[width=0.85\linewidth]{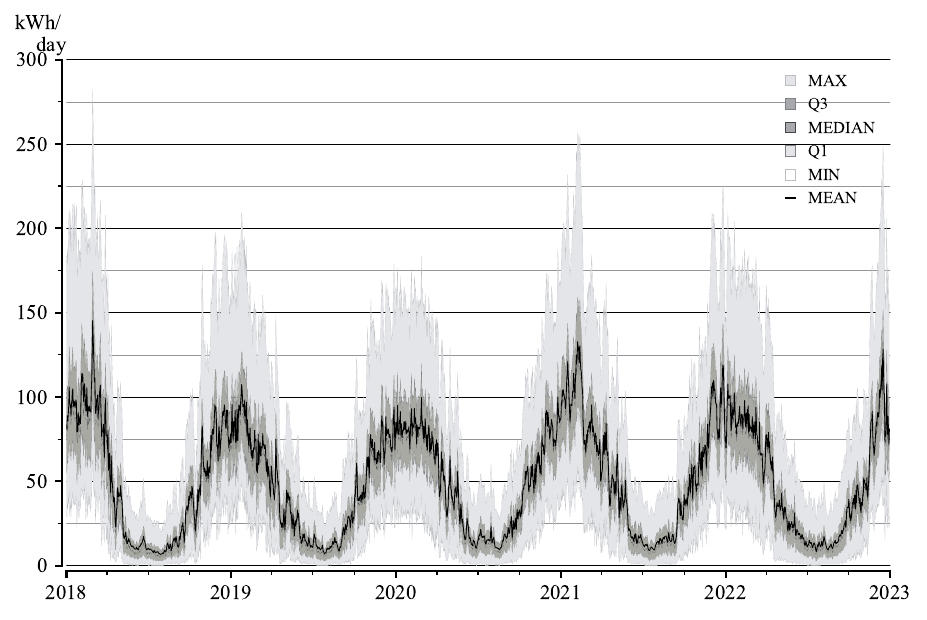}
    \caption{Daily energy use for space heating and DHW across the  Aalborg housing stock - Jan 2018 till Dec 2022}
    \label{fig:method_energy}
\end{figure}

The EPC-labeled dataset cover a wide range of heat consumption. Figure \ref{fig:method_energy} shows the daily energy use for space heating and domestic hot water (DHW) from 2018 to 2023. The median energy use peaks around $100$ kWh/day, with the range of maximum and minimum use during the peak period ranging from 200 to 25 kWh/day. 

\begin{figure}[h]
    \centering
    \includegraphics[width=0.85\linewidth]{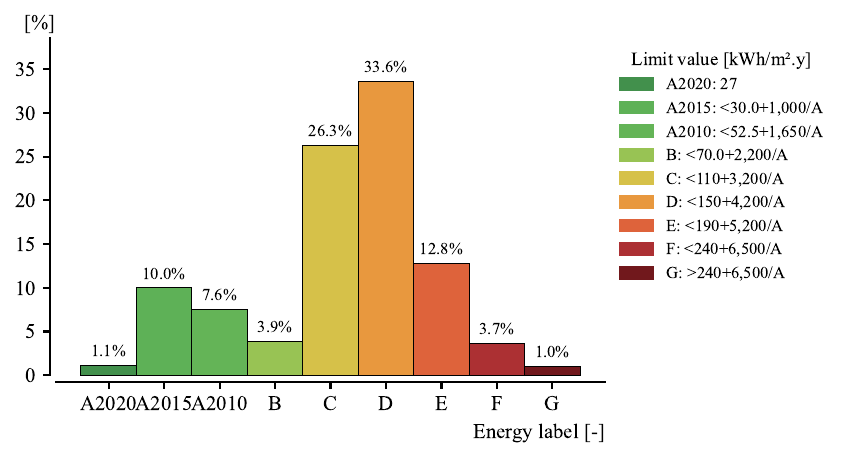}
    \caption{EPC labels Aalborg housing stock}
    \label{fig:method_EPC}
\end{figure}

Another way of examining the coverage of the dataset is though the accompanying EPC labels of the examine single-family homes. Figure \ref{fig:method_EPC} shows the proportion of each EPC label withing the dataset. As can be seen from Figure \ref{fig:method_EPC}, the dataset covers entire range of labels, from A2020 to G. Although the extreme labels such as A2020 and G are both only 1\% of the dataset. Nonetheless, this assessment shows that the selected dataset has a wide coverage of building energy efficiency.\\

\subsubsection{Weather Data}

\noindent As the AAU dataset does not come with accompanying weather data from local weather stations, the weather observations for Aalborg were obtained from the Copernicus Climate Data Store, namely the \textit{ERA5-Land hourly} re-analysis dataset. The dataset is gridded with resolution of $0.1^o \times 0.1^o$ (longitude and latitude) which corresponds to an approximate resolution of 9 km $\times$ 9 km \citep{COPERNICUS}. 

\subsection{Models}

\noindent This section describes the model considered in this investigation, namely the ES model, the ARX-ES model, and the ARMAX-ES models. As the parameter values between the models are not directly comparable, particular care is taken to describe adjustments needed to compare parameter estimates between the ES and ARX-ES models.\\

\subsubsection{Energy Signature (ES) Model - Bayesian Formulation}

\noindent The model described in Section \ref{sec:Compemporary_ES} serves as the base/simplest model against which all extensions are compared. It is a direct translation of the ES model into the Bayesian framework. The complete Bayesian specification of the ES model is provided in Table \ref{table:ES_specification}. \\

\begin{table*}
\centering
\begin{minipage}{0.75\textwidth}
\noindent \textbf{Model Structure:}
\begin{equation}\label{eq:ES_structure}
    \begin{split}
        \mathbf{\Phi}^{(t)}_{\text{ES}} \mid \mu_{(t)}, \sigma_{(t)}  & \sim N(\mu_{(t)}, \sigma_{(t)}^2) \\
        \mu_{(t)} & = \text{LSE}( \mathbf{\Phi}_{\text{Winter}}^{(t)}, \mathbf{\Phi}_{\text{Summer}}^{(t)} ) \\
        \sigma_{(t)} & = \tau^{(t)}\sigma_{\text{reduction}} + (1-\tau^{(t)})\sigma_{\text{winter}}\\
        \mathbf{\Phi}_{\text{Winter}}^{(t)} \mid \text{UA}_{\text{0}}, \text{UA}_{\text{W}}, \text{T}_{\text{b}}, \text{gA} & = (\text{UA}_0 + \text{UA}_{\text{W}}\mathbf{W}_{\text{s}}^{(t)})(\text{T}_{\text{b}} - \mathbf{T}_{\text{a}}^{(t)})\\
        & - \text{gA }\mathbf{I}_{\text{g}}^{(t)} + \Phi_{\text{base}} \\
        \mathbf{\Phi}_{\text{Summer}}^{(t)} \mid \Phi_{\text{base}} & = \Phi_{\text{base}}\\
        \tau^{(t)} & = \frac{\exp{(k\Phi_{\text{Winter}}^{(t)})}}{\exp{(k\Phi_{\text{Winter}}^{(t)})}+\exp{(k\Phi_{\text{Summer}}^{(t)})}}
    \end{split}
\end{equation}

\noindent \textbf{Model Parameter Priors:}

\begin{equation}
    \begin{split}
        \text{UA}_0 & \sim \Gamma(4.75, 0.3) \\
        \text{UA}_{\text{W}} & \sim \Gamma(0.27,0.02) \\
        \text{T}_{\text{base}} & \sim \text{Normal}(18,1.2) \\
        \text{gA} & \sim \Gamma(4, 0.075)  \\
        \log(\sigma_1) & = \Gamma(4.25, 0.82) \\
        \log(\sigma_{\text{reduction}}) & = \Gamma(4.25, 0.82) \\
        k & \sim \Gamma(1, 1)
    \end{split}
\end{equation}
\hrule
\end{minipage}
\label{table:ES_specification}
\caption{Bayesian-ES Model}   
\end{table*}

\subsubsection{Auto-Regressive - Energy Signature (ARX-ES) Model}
\noindent Daily heating loads are expected to show autocorrelation to some extent, influenced by factors such as occupancy and the physical properties of the house. The problem with autocorrelation is that it violates the assumption of independence between data samples, which could lead to biased inference, and uncertainty estimates could be overly optimistic. A way to partly circumvent the issue is to assume autocorrelation in the noise process and add an Auto-Regressive (AR) component to the model.

ARX types of models are being deployed in the building domain \citep{ARX_school} as they still offer a high degree of interpretability while capturing the dynamics. In physical terms, the AR part of the ARX models can be interpreted as capturing thermal inertia in building models \citep{dynamic_horizon}. 

The structure of the ARX-EX model is largely the same as that of ES model, however, interpretation of the parameters requires more care. \\

\noindent First, consider the simple weather independent regime that fluctuates around a level $\Phi_{\text{base}}$ with some persistence determined by $\rho_1$,
\begin{equation}\label{eq:ar_ex_1}
    \mathbf{\Phi}_{\text{Summer}}^{(t)} = \underbrace{\Phi_{\text{base}} + \xi^{(t)}}_{\text{ES-part}} + \underbrace{\rho_1 \mathbf{\Phi}_{\text{Summer}}^{(t-1)}}_{\text{new AR-part}} 
\end{equation}
where $\xi^{(t)} \sim N(0, \sigma^2)$ as previously and $|\rho_1|<1$. Note that this is simply the weather-independent regime from equation \ref{eq:ES_structure} with an added AR(1) component as illustrated by the under-braces of equation \ref{eq:ar_ex_1}.\\

\noindent By use of the relation $\mathbf{\Phi}_{\text{Summer}}^{(t)} = \Delta \mathbf{\Phi}_{\text{Summer}}^{(t)} + \mathbf{\Phi}_{\text{Summer}}^{(t-1)}$, the equation \ref{eq:ar_ex_1} can be reformulated as,
\begin{equation}
    \begin{split}
        \Delta \mathbf{\Phi}_{\text{Summer}}^{(t)} &= \Phi_{\text{base}} + \rho_1 \mathbf{\Phi}_{\text{Summer}}^{(t-1)} - \mathbf{\Phi}_{\text{Summer}}^{(t-1)} + \xi^{(t)}\\
        &= -(1-\rho_1)( \mathbf{\Phi}_{\text{Summer}}^{(t-1)} - \frac{\Phi_{\text{base}}}{(1-\rho_1)}) + \xi^{(t)}\\
        &=-\gamma_2 (  \mathbf{\Phi}_{\text{Summer}}^{(t-1)} - \Phi_{\text{base}}^{\text{long-term}})
    \end{split}
\end{equation}
where $\gamma_2 = 1 - \rho_1$ and $\Phi_{\text{base}}^{\text{long-term}} = \frac{\Phi_{\text{base}}}{(1-\rho_1)}$.\\

\noindent This reparameterisation transforms the model into the error-correction model (ECM), enhancing interpretability. The interpretation being that if $\mathbf{\Phi}_{\text{Summer}}^{(t-1)}$ deviates from the long-term base heat-load $\Phi_{\text{base}}^{\text{long-term}}$, a correction is applied, and the magnitude of the correction is controlled by the size of $\gamma_2$. \\

\noindent The model can also be re-contextualised in another form. Notice that by introducing the back-shift operator B (i.e. $\text{B}\mathbf{\Phi}_{\text{Summer}}^{(t)} = \mathbf{\Phi}_{\text{Summer}}^{(t-1)}$), the model can be re-written as:
\begin{equation}
    (1-\rho_1 \text{B}) \mathbf{\Phi}_{\text{Summer}}^{(t)} = \Phi_{\text{base}} + \xi^{(t)}
\end{equation}
this can also be rewritten in the transfer-function form which yields
\begin{equation}
    \mathbf{\Phi}_{\text{Summer}}^{(t)} = \frac{1}{1-\rho_1} \Phi_{\text{base}} + \frac{1}{1-\rho_1} \xi^{(t)}
\end{equation}
This is also known as the steady-state transformation, and once again, $\Phi_{\text{base}}^{\text{long-term}} = \frac{\Phi_{\text{base}}}{(1-\rho_1)}$ appears, corresponding to the long-term base heat-load from the ECM formulation, emphasizing the shared concepts between the domains. 

It is these long-term parameters that should be compared with the parameters found in the static models like the earlier ES model.\\

\noindent Now, consider the weather-dependent regime in \ref{eq:ar_ex_2}. 

\begin{table*}[h]
\centering
\begin{equation}\label{eq:ar_ex_2}
    \begin{split}
        \mathbf{\Phi}_{\text{Winter}}^{(t)} &= \underbrace{(\text{UA}_0 + \text{UA}_{\text{W}}\mathbf{W}_{\text{s}}^{(t)})(\text{T}_{\text{b}} - \mathbf{T}_{\text{a}}^{(t)}) - \text{gA}\mathbf{I}_{\text{g}}^{(t)} + \Phi_{\text{base}} + \xi^{(t)}}_{\text{old ES-part}}\\
        &+ \underbrace{\rho_1 \mathbf{\Phi}_{\text{Winter}}^{(t-1)}}_{\text{new AR-part}}\\
        &= \underbrace{\text{UA}_0 \text{T}_{\text{b}} + \text{UA}_{\text{W}} \text{T}_{\text{b}} \mathbf{W}_{\text{s}}^{(t)} + \text{UA}_0 \mathbf{T}_{\text{a}}^{(t)} + \text{UA}_{\text{W}} \mathbf{T}_{\text{a}}^{(t)} \text{W}_{\text{s}}^{(t)}}_{\text{Expanded temperature dependent component}}\\
        & - \text{gA}\mathbf{I}_{\text{g}}^{(t)} + \Phi_{\text{base}} + \xi^{(t)}\\
        &+ \rho_1 \mathbf{\Phi}_{\text{Winter}}^{(t-1)}\\
        &= \text{x}^{(t)} \theta + \Phi_{\text{base}} + \xi^{(t)} + \rho_1 \mathbf{\Phi}_{\text{Winter}}^{(t-1)}
    \end{split}
\end{equation}

\end{table*}

where $\theta = [\text{UA}_0 \text{T}_{\text{b}};\text{UA}_{\text{W}} \text{T}_{\text{b}}; \text{UA}_0; \text{UA}_{\text{W}}]$ and $\text{x}^{(t)} = [1; \mathbf{W}_{\text{s}}^{(t)}; \mathbf{T}_{\text{a}}^{(t)}; \mathbf{T}_{\text{a}}^{(t)}\mathbf{W}_{\text{s}}^{(t)}]$.\\

\noindent This time consider also using the relation $ x^{(t)} = \Delta x^{(t)} + x^{(t-1)}$ and apply it to \ref{eq:ar_ex_2} to obtain. 

\begin{equation}\label{eq:ARX_ES_short_long}
    \begin{split}
        \Delta \mathbf{\Phi}_{\text{Winter}}^{(t)} &= \Phi_{\text{base}} + (\Delta x^{(t)} + x^{(t-1)})\theta\\
        & - (1 - \rho) \mathbf{\Phi}_{\text{Winter}}^{(t-1)} + \xi^{(t)}\\
        &= \Phi_{\text{base}} + (\Delta x^{(t)} + x^{(t-1)})\theta\\
        &- \gamma \mathbf{\Phi}_{\text{Winter}}^{(t-1)} + \xi^{(t)}\\
        &= \underbrace{\Delta x^{(t)}\theta}_{\text{short-term}}\\ 
        &- \gamma \underbrace{(\mathbf{\Phi}_{\text{Winter}}^{(t-1)} - (\Phi_{\text{base}}^{\text{long-term}} + x^{(t-1)}\theta^{\text{long-term}}))}_{\text{long-term effects}}\\ 
        &+ \xi^{(t)}
    \end{split}
\end{equation}

\noindent where $\gamma = 1 - \rho$, $\Phi_{\text{base}}^{\text{long-term}} = \frac{\Phi_{\text{base}}}{\gamma}$, and $\theta^{\text{lt}} = \frac{\theta}{\gamma}$.\\

\noindent Notice that $\text{UA}_0^{\text{long-term}}$, $\text{UA}_{\text{W}}^{\text{long-term}}$, $\text{gA}^{\text{long-term}}$ and $\text{T}_{\text{b}}^{\text{long-term}}$ are contained within $\theta^{\text{long-term}}$ and can be used to compare models without dynamic specification. Although the parameters contained within $\theta^{\text{long-term}}$ are of primary importance, the parameters contained in $\theta$ can still be informative and should be considered as short-term effects of the meteorological variables with immediate impact on the heat-load.

The explicit formulation of the Bayesian ARX-ES model is presented in Table \ref{table:ARX_ES_specification}.\\

\subsubsection{Auto-Regressive Moving-Average - Energy Signature (ARMAX-ES) Model}

\noindent The last model considered in this paper is the ARMAX-ES model, which is the Auto-Regressive Moving Average with eXogeneous input (ARMAX) model adapted to the Energy Signature model. In essence, this model is the natural extension of the ARX-ES model after inclusion of the Moving-Average (MA) component.\\

\noindent Using the same conceptualization as in equation \ref{eq:ARX_ES_short_long}, notice that the same model interpretation applies to the ARMAX-ES model, and only the parameters from $\theta^{\text{long-term}}$ can be directly compared with non-dynamic models such as the ES model. While the MA term does not directly influence interpretation of parameters in $\theta^{\text{long-term}}$, the MA parameters have additional interpretation of their own. The ARMAX models can be shown to be a discrete time approximation of the popular Grey-Box RC models, which are used to describe heating dynamics with stochastic differential equations \citep{ARMAX_physical}. Bringing the models ARMAX models close to state-of-art physics based Grey-Box models.

Lastly, the ARMAX models were implemented via the methology suggested by H.Spliid \citep{Spliid01121983} within the \texttt{PyMC} framework.

\section{Results}\label{sec:Results}

\noindent This section presents evaluation of the Bayesian Energy Signature (ES) models and their extensions, including the ARX-ES and ARMAX-ES models, applied on district heating data of Danish single-family homes (n=2788). The analysis focuses on model validation, predictive performance, and parameter estimation, highlighting the added value of incorporating autoregressive and moving-average components. Key metrics such as Bayesian $R^2$, Expected Log Pointwise Predictive Density (ELPD), and posterior distributions are examined to assess model accuracy and reliability. In addition, a single-building case study demonstrates the practical implications of Bayesian uncertainty quantification and its utility in real-world energy diagnostics. Together, these results illustrate the strengths and trade-offs of the proposed Bayesian framework for building energy performance modeling.

\subsection{Performance and Validation}

\subsubsection{Model Validation}

\noindent Before examining the parameter estimates of the building population, and model inference on the individual buildings, it is necessary to validate and compare model performance. Begin by examining in-sample and out-of-sample model performance. 

\begin{figure}[h]
    \centering
    \includegraphics[width=0.85\linewidth]{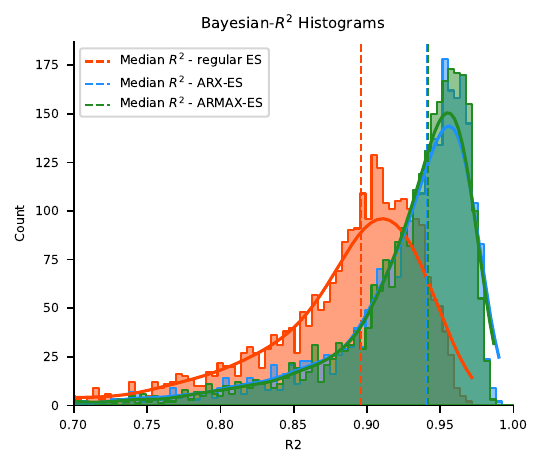}
    \caption{Histograms of mean Bayesian $R^2$ performance metrics for Bayes-ES, ARX-ES and ARMAX-ES models (red, blue and green respectively).}
    \label{fig:Results_R2}
\end{figure}

The histograms of the mean Bayesian $R^2$ score for the ES, ARX-ES and ARMAX models in Figure \ref{fig:Results_R2}, clearly shows that the performance is substantially improved with increased model complexity. The median model performance is improved from approx $R^2_{\text{ES}}=0.89$ to $R^2_{\text{ARMAX-ES}}=0.94$. The entire distribution of the ARMAX-ES model is shifted closer towards 1 compared to the simpler models. The improvement from ARX to ARMAX models is much less pronounced. The improvement can be most readily seen from more concentrated peak of the histograms where the ARMAX-ES model is far more concentrated.

\begin{figure}[h]
    \centering
    \includegraphics[width=0.85\linewidth]{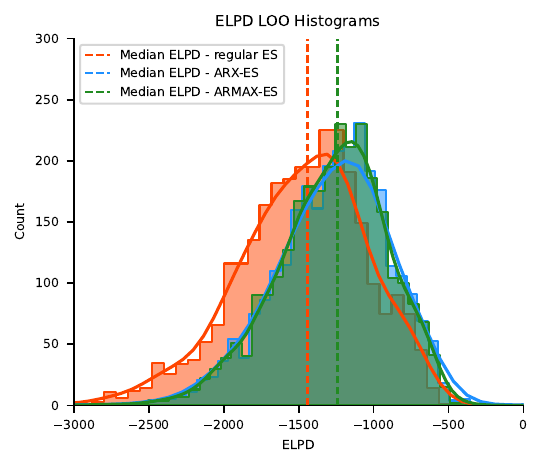}
    \caption{ELPD Performance metrics for Bayes-ES, ARX-ES and ARMAX-ES models.}
    \label{fig:Results_ELPD}
\end{figure}

\noindent Same observations can be inferred from examining the histograms of $\text{ELPD}_{\text{LOO}}$ score in Figure \ref{fig:Results_ELPD}, corresponding to the out-of-sample performance estimate. The figure shows three histograms which are also shifting rightwards with increasing model complexity. Once again, the difference between the ARX and ARMAX scores appears to be minor.\\

\begin{figure*}[h]
    \centering
    \includegraphics[width=0.95\linewidth]{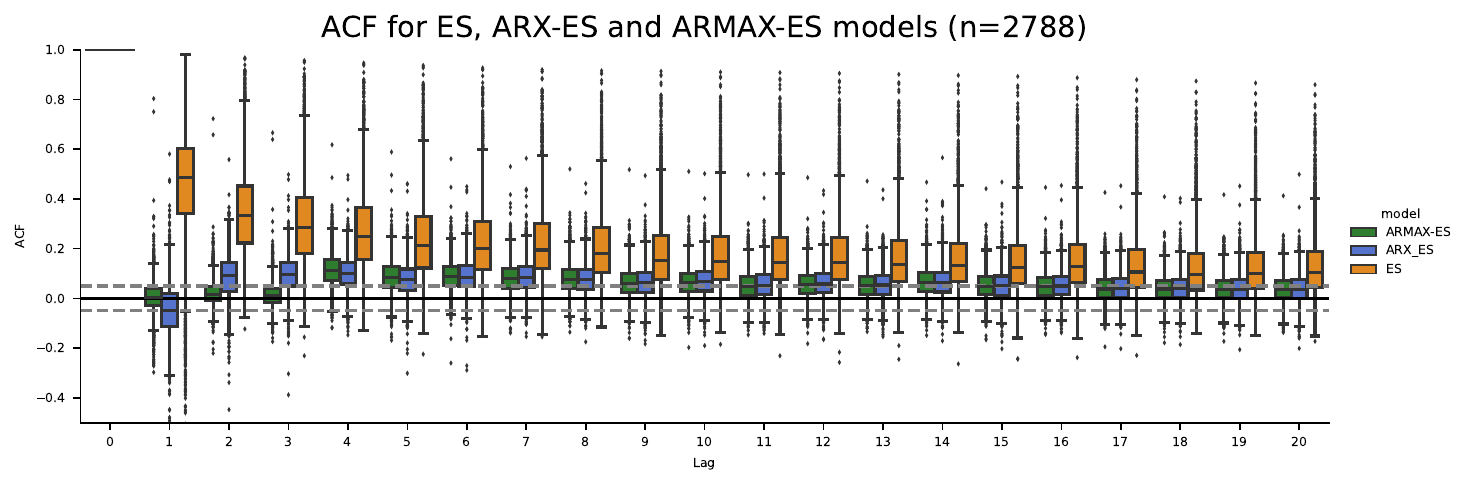}
    \caption{Box plot of ACF coefficient for ES, ARX-ES and ARMAX-ES models (n=2877) for 20 lags. Closer to zero is better.}
    \label{fig:ACF_boxplot}
\end{figure*}

\noindent Next, examine the ACF computed on residuals on each building in the dataset. The ACF estimates for the building population are visualized in Figure \ref{fig:ACF_boxplot} as box plots for the first 21 lags for the ES, ARX and ARMAX model. The ACF function, one of key metrics in time-series analysis, denotes amount of auto-correlation between residual observations, which is undesirable. Unsurprisingly, the ES model presents a high degree of auto-correlation, with box-plot for each lag showing that a substantial amount of observations above 0.1. The largest improvement is achieved in transition from ES to the ARX model, with marginal improvements from ARMAX-ES model, which are mainly seen through decreased spread of ACF components. It must be noted that while the mean ACF for ARMAX-ES models are close or below 0.05, in general, a small degree of auto-correlation is still present in the residuals. While the improvements ARX and ARMAX models are minor, it is important to examine how they relate to model parameters estimated by each model.\\

\subsection{Building Population Posterior Parameter Distributions}

\begin{figure*}[h]
    \centering
    \includegraphics[width=0.9\linewidth]{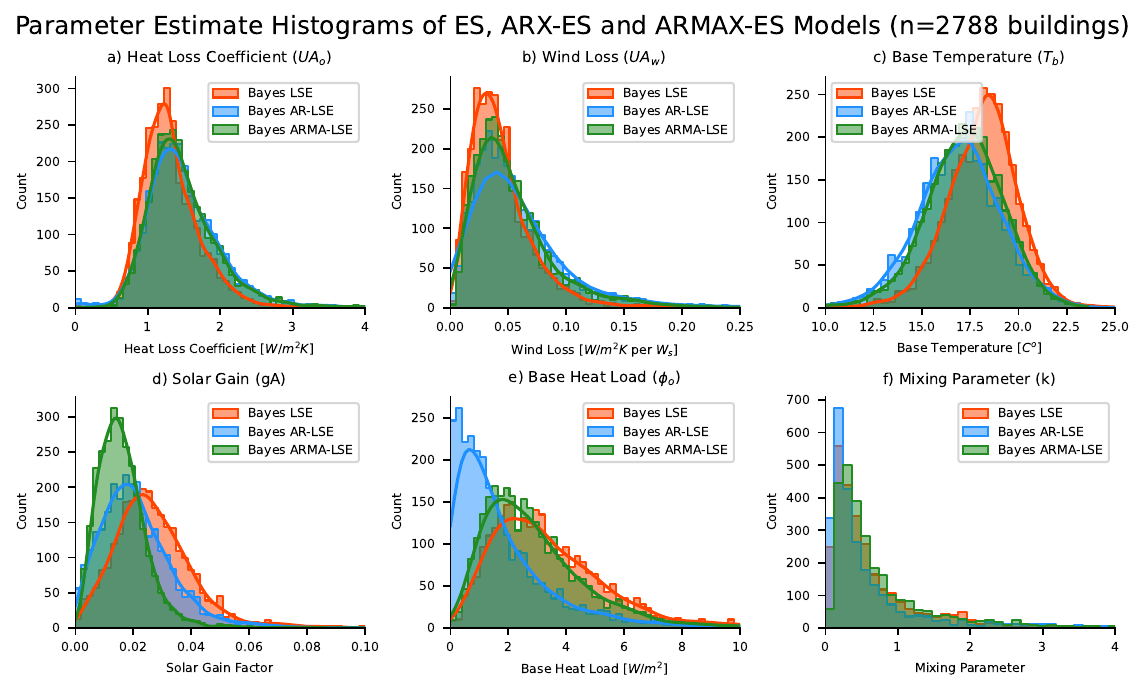}
    \caption{Posterior parameter distribution of common model parameters between LSE, AR-LSE and ARMA-LSE models.}
    \label{fig:Results_Par_Dist}
\end{figure*}

\noindent Examining the parameter distribution of the three fitted models, namely ES, ARX-ES and ARMAX-ES one would expect that the mean parameter estimates would be the same. Figure \ref{fig:Results_Par_Dist} shows that this is not always the case. The histograms of the mean estimated model parameters largely coincide for the three models, and all histograms appear to be vaguely gamma-shaped, apart from the distribution of $\text{T}_b$ parameter, which is symmetric. One instance where the distributions notably do not coincide is the Base Heat Load ($\Phi_{\text{base}}$) parameter. While the distribution for $\Phi_{\text{base}}$ between the ES and ARMAX-ES are nearly identical, the distribution for ARX-ES is considerably shifted towards zero. Additionally, notice that only ARX estimates for HLC reach zero, which is highly unlikely. This is highly suggestive of bias within ARX model parameter estimates, which is later mitigated by additional MA component in ARMAX-ES model. 

The parameter estimates of most interest is the Heat Loss Coefficient and Wind Penetration, as they are among leading causes of building's heat loss. As seen in Figure \ref{fig:Results_Par_Dist}, these parameters are also the ones where the histograms between the three models largely coincide. The HLC estimates for the building population range from 0.5 to 4 [$\text{W}/\text{m}^2\text{K}$] with median value of 1.4 [$\text{W}/\text{m}^2\text{K}$] where the distribution is not symmetric but skewed to the left. For the wind penetration coefficient ($\text{UA}_{\text{W}}$), the values range between 0 and 0.25 with median of 0.045 [$\text{W}/\text{m}^2\text{K}$ per $\text{m}/\text{s}$ (Wind Speed)] with the distribution being even further skewed leftwards towards zero.

\subsection{Single-Building Results Showcase}

\noindent While the population-level parameter estimate results are interesting, their accuracy is dependent on the accuracy of the individual building models. For this, examine the results of a single representative building. Evaluation of individual buildings is also the context in which the uncertainty quantification of the Bayesian methodology can be demonstrated.

\begin{figure*}[h]
    \centering
    \includegraphics[width=0.95\linewidth]{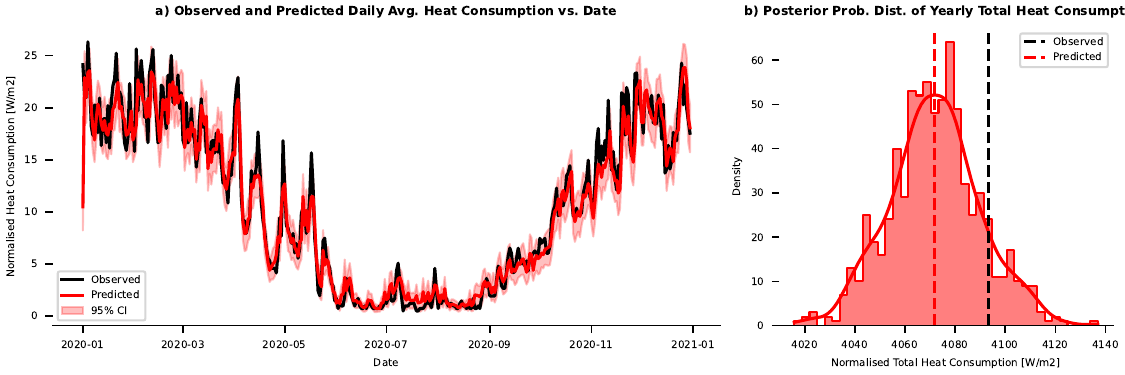}
    \caption{a) Observed and Predicted Daily Avg. Heat Consumption. b) Posterior Predictive Distribution of normalised annual heat consumption for example building.}
    \label{fig:Results_Single_Yearly}
\end{figure*}

Figure \ref{fig:Results_Single_Yearly} shows the predicted energy consumption for a single building against actual consumption. The figure shows that the ARMAX-ES model is able to capture the seasonal variation in energy consumption. The model is able to predict the energy consumption with a high degree of accuracy, with the predicted energy consumption closely following the actual consumption. 

Furthermore, examining the posterior probability distribution for total yearly consumption, it is clear that the model is able to accurately estimate the total yearly consumption. The distribution is contains the actual consumption close to the center, with a small spread. The mean of predicted posterior is only $0.05\%$ off from actual observed yearly consumption, and the 95\% quantiles of the distribution are only $\pm 1 \%$ from the mean.\\

\begin{figure}[h]
    \centering
    \includegraphics[width=0.85\linewidth]{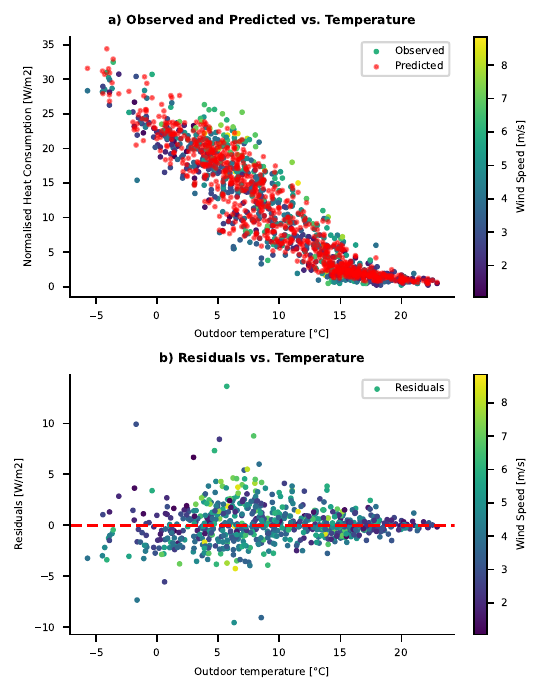}
    \caption{\textbf{a)} Predicted heat load profile for a single building in red, with actual heat load in the background against outdoor temperature. \textbf{b)} Residuals against outdoor temperature, colored by wind-speed.}
    \label{fig:Results_ES_Profile}
\end{figure}

\noindent The predicted heat load can also be examined in the typical ES plot, where the heat load is plotted against the outdoor temperature. Figure \ref{fig:Results_ES_Profile} shows the heat load profile for the same building. The figure shows that the ARMAX-ES model is able to accurately capture the heat load profile. Note that the predicted heat load covers the spear of the actual heat load. Furthermore, examining the residuals, it is clear that the residuals are mean-zero with not clear pattern based on wind-speed.\\

\begin{figure*}[h]
    \centering
    \includegraphics[width=0.8\linewidth]{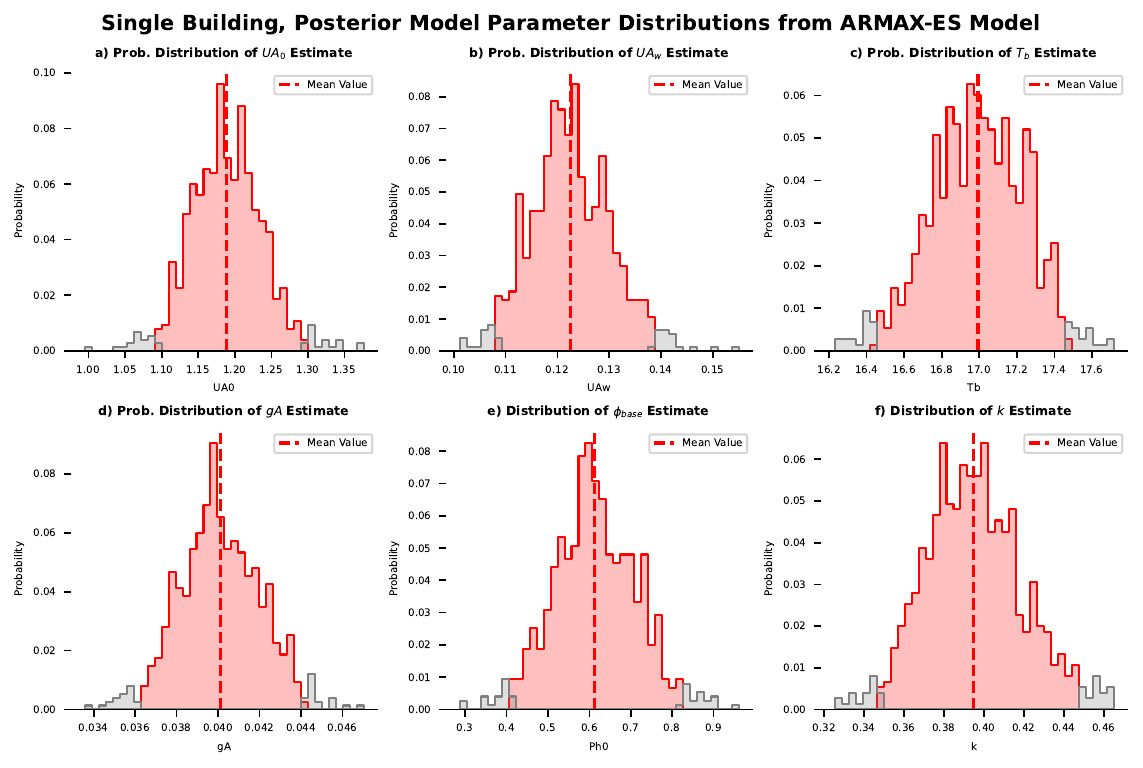}
    \caption{Posterior distributions of the estimated model parameters for the example building. The red area corresponds to the 95\% quantile.}
    \label{fig:Results_Single_Building_Posterior}
\end{figure*}

\begin{figure*}[h]
    \centering
    \includegraphics[width=0.8\linewidth]{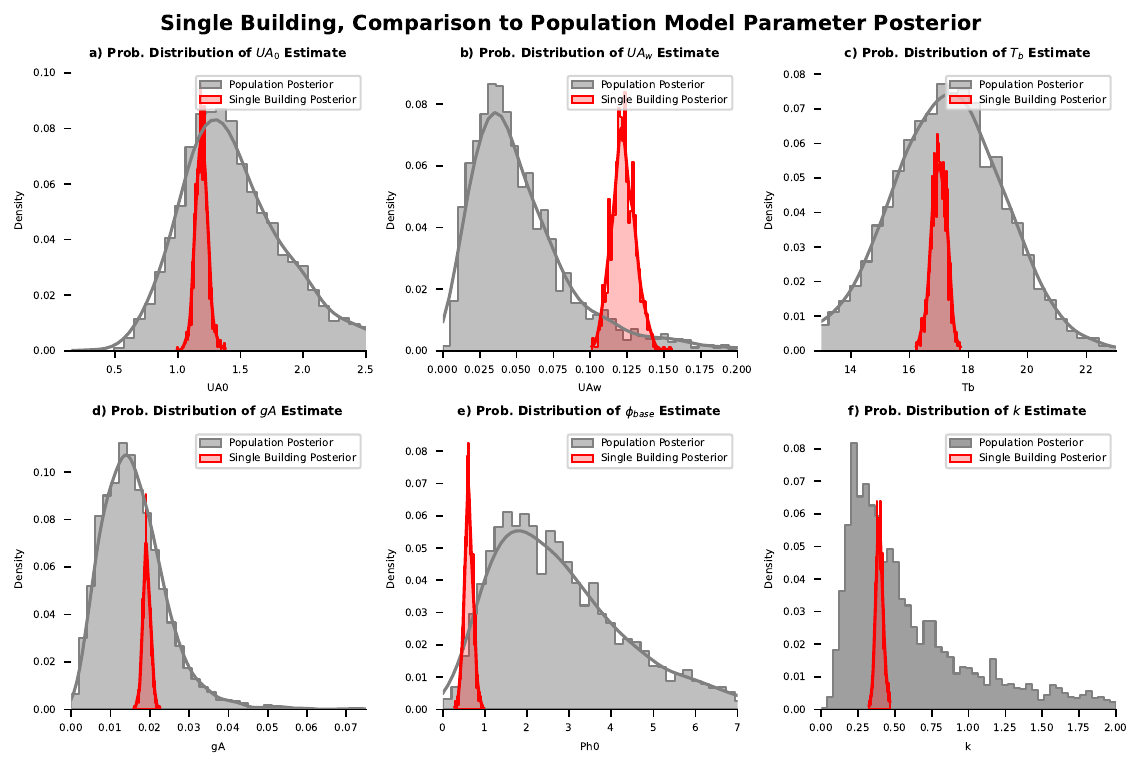}
    \caption{Posterior distributions of the estimated model parameters for the example building (red) compared to distribution of parameters for all buildings in the dataset (grey).}
    \label{fig:Results_Posterior_Comparison_To_Population}
\end{figure*}

\noindent Lastly, examine the posterior distribution of the parameters for the single building in Figure \ref{fig:Results_Single_Building_Posterior}. From the figure, it is clear that the posterior histograms of each parameter are well-behaved, symetrical, and vaguely normal-shaped. However, note that the 95\% quantiles for the parameter estimates have significant width, e.g. in case of the $\text{UA}_0$ estimate, the quantiles correspond to around $\pm8\%$ spread around the mean, describing the inherent uncertainty in the parameter estimates based on the given data.

However, these model parameter posterior should also be examined in the context of entire building population. For this, see Figure \ref{fig:Results_Posterior_Comparison_To_Population}, where it can be seen that the parameter estimates and their CIs represent a fraction of the range of possible population parameter estimates and informative comparison can be made between the single building and the population. E.g. the $\text{UA}_0$ estimate for the single building is within the median range of the population estimates, but the $\text{UA}_w$ estimate is at the far lower end of the population estimates, while the base heat-load parameter $\Phi_{\text{base}}$ parameter is very low compared to the entire population.\\
\section{Discussion}\label{sec:Discussion}

\noindent This section first presents an evaluation of the model's performance and validation, analyzing key predictive performance metrics, inter-model parameter comparisons, and insights from single-building analyses. It then discusses the broader implications of the study, focusing on the practical benefits of Bayesian modeling for different stakeholders, including building owners, engineers, and energy providers. Finally, the section addresses the limitations of this study, highlighting challenges related to model scalability, computational demands, and practical applicability in diverse building engineering contexts.

\subsection{Model Performance and Validation}

\subsubsection*{Predictive Performance Considerations}

\noindent In terms of one-step prediction, the model's performance metrics—including $\text{R}^2_{\text{Bayes}}$, Expected Log Pointwise Predictive Density (ELPD), and Autocorrelation Function (ACF)—improve with increasing model complexity (see Figure \ref{fig:Results_R2}). The highest one-step prediction performance is achieved with the most complex ARMAX-ES model, which demonstrates the ability to capture a greater amount of variability in the data. This result suggests that ARMAX-ES can effectively leverage all available information from daily building data. Note that the main goal of this methodology is to maximize description of the data by the model, and not create a generalized model that performs well on unseen data, hence a lot of emphasis is placed on in-sample $R^2$ metric and parameter posteriors.

However, it is important to recognize that this short-term predictive accuracy does not necessarily translate to the long-term forecasting required for estimating total annual heat demand. The ARMAX-ES model’s high one-step accuracy primarily reflects its capacity to make use of short-term data dependencies rather than providing superior long-term forecasts. In contrast, the basic Energy Signature (ES) model proves adequate for long-term forecasts, particularly when estimating annual heat demand. The Law of Large Numbers favors the ES model over long horizons, supported by the fact that residuals tend to average out over time, making it a reliable choice for yearly heat-demand calculations.\\

\subsubsection*{Inter-Model Parameter Estimate Comparison}

\noindent When comparing parameter estimates presented in Figure \ref{fig:Results_Par_Dist}, it is clear that the parameters estimated between the models are generally aligned, however the estimated from the ARX-ES model stand out. While the ARX-ES model achieves improved predictive performance compared to the ES model, and is comparable to the ARMAX-ES (see Figure \ref{fig:Results_R2}), its parameter estimates can diverge significantly from those obtained by the ES and ARMAX-ES model. Two such cases are base heat-load and heat-loss coefficient for very low values (see Figure \ref{fig:Results_Par_Dist}) where the values can become unrealistically close to zero. This suggests that while the ARX-ES model is effective for simpler, short-term forecasts, it may introduce bias due to an over-reliance on auto-regressive observations rather than exogenous inputs. The performance of the ARMAX-ES model remains superior, but for tasks requiring simpler estimates of basic building characteristics, the ES model remains the more effective and reliable tool.\\

\subsubsection*{Single-Building Results}

\noindent The single-building analysis demonstrates the true value of the Bayesian approach and the ARMAX-ES model's capability to meaningful insights into building's energy consumption patterns. As shown in Figures \ref{fig:Results_Single_Yearly} and \ref{fig:Results_ES_Profile}, the ARMAX-ES model captures seasonal and temperature-dependent variations with great accuracy, however the most noteworthy feature of the Bayesian approach lies in its uncertainty quantification. For instance, the posterior distribution for total yearly consumption reveals that the predicted mean aligns closely with the observed value, with a minimal deviation of only $0.05\%$. Furthermore, the narrow 95\% credible interval (within $\pm 1\%$ of the mean) highlights the model's precision and reliability in annual energy estimation. This level of accuracy and uncertainty representation is particularly advantageous for stakeholders making investment or operational decisions.

Moreover, the analysis of parameter posteriors (Figure \ref{fig:Results_Single_Building_Posterior}) demonstrates well-behaved distributions with symmetrical and approximately normal shapes, reflecting the model's robustness. However, the credible intervals for parameters such as $\text{UA}_0$ reveal inherent uncertainties, e.g., an 8\% spread around the mean. These quantified uncertainties provide essential insights for diagnostics and decision-making, as they reflect both the limitations of the data and the underlying model assumptions.

When comparing individual building results to population-level distributions (Figure \ref{fig:Results_Posterior_Comparison_To_Population}), the ARMAX-ES model offers a unique perspective. For instance, the $\text{UA}_0$ parameter for the single building falls within the median range of the population, suggesting consistency with broader trends. Conversely, other parameters, such as $\text{UA}_w$ and $\Phi_{\text{base}}$, deviate significantly, emphasizing the value of analyzing both individual and population-level dynamics. This dual insight enables targeted diagnostics and highlights the model's potential for both diagnostics and decision-making in the context of energy performance management.

\subsection{Implications of the study}

\noindent Bayesian methods provide distinct advantages for different stakeholders in the building industry.

\paragraph*{Building Owners}  

For building owners, the Bayesian Energy Signature (ES) approach offers a robust framework for gaining an accurate, data-driven assessment of their building’s energy performance. This methodology not only enables owners to develop a more realistic understanding of their property’s energy demand but also provides clear financial insights, empowering them to make informed decisions. Owners can identify specific areas where upgrades may yield the highest returns, establishing a stronger business case for investments in energy-saving improvements. The probabilistic estimates generated by Bayesian methods further aid in anticipating variability in annual heating costs, allowing owners to budget more effectively and reduce exposure to energy price fluctuations.

\paragraph*{Building Engineers}

Bayesian methods provide engineers with a tool to refine their understanding of a building’s thermal characteristics by integrating real, variable data and expert knowledge directly into the modeling process. This approach improves the reliability of the heat-loss coefficient (HLC) and other key parameters, offering engineers an opportunity to adopt more data-driven methodologies in EPC certification. Ultimately, these insights can contribute to revising the established building knowledge base, fostering a shift toward more dynamic, empirically grounded building assessments that reflect the in-use performance of buildings.

\paragraph*{Energy Providers}

Energy providers can benefit from a comprehensive quantification of uncertainty across the building stock they service. By having a more detailed understanding of the performance of buildings, energy providers can better anticipate demand variability, thereby improving their capacity for load planning and resource allocation \citep{DU2025134934, LEONCINI2025134522, KAUKO2025135310}. This knowledge is particularly valuable in preparing for extreme weather events or shifts in user behavior, allowing providers to implement more accurate and resilient energy strategies.

\subsection{Limitations of the study}

\noindent While this research demonstrates the efficacy of Bayesian models in assessing building energy performance, several practical constraints should be acknowledged.

\paragraph{Model Application on Apartment and Commercial Buildings}

The main focus of this study was on single-family houses, which represent a significant portion of the building stock. However, the application of Bayesian models to apartment buildings and commercial properties may present additional challenges. These buildings often have more complex energy systems, such as shared heating systems or variable occupancy patterns, which can complicate the modeling process. One immediate extension that can be added to account for these is of a periodic spline component which could track weekly trends, as was done in \citep{Justinas-Splines}. However, future research should explore the adaptation of Bayesian methods to these building types, considering the unique characteristics and data availability of each.

\paragraph{Computational Intensity Trade-offs}

While Bayesian methods offer significant advantages in terms of flexibility and uncertainty quantification, they are not without drawbacks. One of the most significant challenges is their computational intensity. \textit{Unlike frequentist methods, which typically involve straightforward optimization techniques, Bayesian models require sampling from the posterior distribution.} The posterior distribution most often can not be expressed analytically, and thus has to be sampled using numerical methods such as Markov Chain Monte Carlo (MCMC), which can be computationally expensive, especially when dealing with large datasets or complex models.

In the context of building energy modelling, where high-frequency data (e.g., from smart meters) may be available, the computational burden can be a limiting factor. To mitigate this, in the present study, the models are fitted to daily-average heat demand data. This temporal resolution captures key building characteristics while reducing the computational complexity, and the results can be used to inform more detailed Frequentist forecasting models based on hourly data.

\paragraph*{Model Complexity and Practical Application}
The mathematical complexity of Bayesian models can be a barrier to widespread adoption among practitioners who may lack advanced statistical knowledge. Although modern tools such as PyMC \citep{abril2023pymc} have made Bayesian modeling more accessible, the setup and interpretation of results still require a solid understanding of probabilistic modeling and statistical inference, posing a potential barrier to engineers trained primarily in deterministic approaches.

\paragraph*{Actionability in Real-World Scenarios}
While Bayesian models provide valuable insights into building performance, translating these insights into actionable recommendations for real-world applications remains challenging. Effective implementation often requires comprehensive knowledge of both building performance and business considerations. For instance, while the models can indicate potential performance improvements, assessing the feasibility and cost-effectiveness of specific interventions—such as insulation upgrades or window replacements—demands additional expertise. Moreover, adapting this approach to buildings with more complex energy systems, such as those with heat pumps or dual heating and cooling requirements, requires further model customization.

\paragraph*{Limitations in Data Specificity}
This study utilized satellite-imputed weather data, which, while generally accurate, may lack the resolution to capture microclimatic variations in urban areas or specific building sites. Access to localized weather data from nearby stations could enhance model precision, especially in regions with highly variable weather patterns.
\section{Conclusion}\label{sec:Conclussion}

\noindent This study presents a novel Bayesian framework for assessing building energy performance using Energy Signature (ES) models by demonstrating it on a dataset of Danish single-family homes (n=2788). By incorporating Bayesian principles, this research offers robust methods for quantifying uncertainty, providing stakeholders with actionable insights that extend beyond the capabilities of traditional deterministic approaches. The findings highlight the advantages of combining Bayesian modeling with contemporary ES extensions, such as ARX and ARMAX components, to improve both predictive accuracy and parameter reliability.

Key results demonstrate that the ARMAX-ES model achieves the highest predictive performance in terms of median $R^2=0.94$, ELPD, and residual ACF metrics, effectively capturing short-term variations in energy demand. However, simpler ES models remain reliable tools for long-term energy forecasting, emphasizing their practicality for estimating annual heat demand. At the individual building level, the Bayesian approach's ability to quantify uncertainty offers unique diagnostic capabilities, enabling stakeholders to make informed decisions tailored to specific building characteristics and performance patterns.

The study also underscores the scalability of Bayesian ES models for broader building assessments, from single-family homes to potentially more complex commercial or apartment properties. While computational intensity and model complexity pose challenges, the use of daily-average data mitigates some of these issues, offering a pathway for practical implementation.

The implications of this research are far-reaching. Building owners gain access to probabilistic estimates that support financial planning and targeted energy upgrades. Engineers are equipped with tools to refine energy performance metrics and challenge traditional EPC assumptions, fostering a shift toward dynamic, data-driven assessments. For energy providers, detailed uncertainty quantification across building stocks enhances resource allocation and demand forecasting.

Despite its contributions, this study acknowledges several limitations, including challenges in adapting the models for complex energy systems and reliance on satellite-imputed weather data. Future work should address these limitations by exploring finer temporal resolutions, incorporating localized weather data, and extending applications to diverse building types and energy systems.

In conclusion, Bayesian ES models represent a powerful advancement in building energy performance evaluation, bridging the gap between empirical data analysis and practical energy management. By enabling more precise and reliable insights, this approach supports the transition to energy-efficient and decarbonized building stocks, aligning with global sustainability goals.

\subsection{Special Thanks}

\noindent This research was made possible through \href{https://www.digitalenergyhub.com}{Digital Energy Hub} and support from \href{https://www.centerdenmark.com}{Center Denmark}.

% Generated by IEEEtran.bst, version: 1.14 (2015/08/26)

\newpage
\newpage
\section{Supplementary Material}

*This contains a number of sections that are relevant as supplementary material to the manuscript, yet are outside the scope of the main text for reasons of limiting word count, improving text flow, expected prior knowledge, and effective writing.

\subsection{Bayesian ARX-ES and ARMAX-ES model specification}

\begin{table*}
    \caption{Bayesian-ARX-ES Model}   
\centering
\begin{minipage}{0.75\textwidth}
\noindent \textbf{Model Structure:}

\begin{equation}
    \begin{split}
        \Phi^{(t)}_{\text{ES}} \mid \mu_{(t)}, \sigma_{(t)}  & \sim N(\mu_{(t)}, \sigma_{(t)}^2) \\
        \mu_{(t)} & = \text{LSE}s( \Phi_{\text{Winter}}^{(t)}, \Phi_{\text{Summer}}^{(t)} ) \\
        \sigma_{(t)} & = \tau^{(t)}\sigma_{\text{reduction}} + (1-\tau^{(t)})\sigma_{\text{winter}}\\
        \Phi_{\text{Winter}}^{(t)} \mid \text{UA}_{\text{0}}, \text{UA}_{\text{W}}, \text{T}_{\text{b}}, \text{gA} & = (\text{UA}_0 + \text{UA}_{\text{W}}\text{W}_{\text{s}}^{(t)})(\text{T}_{\text{b}} - \text{T}_{\text{a}}^{(t)}) - \text{gA}\text{I}_{\text{g}}^{(t)}\\
        &+ \Phi_{\text{base}} + \rho_1 \Phi_{\text{Winter}}^{(t-1)} \\
        \Phi_{\text{Summer}}^{(t)} \mid \Phi_{\text{base}} & = \Phi_{\text{base}} + \rho_1 \Phi_{\text{Summer}}^{(t-1)} \\
        \tau^{(t)} & = \frac{\exp{(k\Phi_{\text{Winter}}^{(t)})}}{\exp{(k\Phi_{\text{Winter}}^{(t)})}+\exp{(k\Phi_{\text{Summer}}^{(t)})}}
    \end{split}
\end{equation}

\noindent \textbf{Model Parameters:}

\begin{equation}
    \begin{split}
        \text{UA}_0 & \sim \Gamma(4.75, 0.3) \\
        \text{UA}_{\text{W}} & \sim \Gamma(0.27,0.02) \\
        \text{T}_{\text{base}} & \sim \text{Normal}(18,1.2) \\
        \text{gA} & \sim \Gamma(4, 0.075)  \\
        \sigma_{\text{winter}} & = \Gamma(4.25, 0.82) \\
        \sigma_{\text{reduction}} & = \Gamma(4.25, 0.82) \\
        k & \sim \Gamma(1, 1)
    \end{split}
\end{equation}
\hrule
\end{minipage}
\label{table:ARX_ES_specification}
\end{table*}

\begin{table*}
\caption{Bayesian-ARMAX-ES Model}   
\centering
\begin{minipage}{0.75\textwidth}
\noindent \textbf{Model Structure:}

\begin{equation}\label{eq:ARMAX_ES_structure}
    \begin{split}
        \Phi^{(t)}_{\text{ES}} \mid \mu_{(t)}, \sigma_{(t)}  & \sim N(\mu_{(t)}, \sigma_{(t)}^2) \\
        \mu_{(t)} & = \text{LSE}s( \Phi_{\text{Winter}}^{(t)}, \Phi_{\text{Summer}}^{(t)} ) \\
        \sigma_{(t)} & = \tau^{(t)}\sigma_{\text{reduction}} + (1-\tau^{(t)})\sigma_{\text{winter}}\\
        \Phi_{\text{Winter}}^{(t)} \mid \text{UA}_{\text{0}}, \text{UA}_{\text{W}}, \text{T}_{\text{b}}, \text{gA} & = (\text{UA}_0 + \text{UA}_{\text{W}}\text{W}_{\text{s}}^{(t)})(\text{T}_{\text{b}} - \text{T}_{\text{a}}^{(t)}) - \text{gA}\text{I}_{\text{g}}^{(t)}\\
        &+ \Phi_{\text{base}} + \rho_1 \Phi_{\text{Winter}}^{(t-1)} + \underbrace{\sum_{i=1}^3 \nu_i e^{t-i}}_{\text{new MA-term}}  \\
        \Phi_{\text{Summer}}^{(t)} \mid \Phi_{\text{base}} & = \Phi_{\text{base}}  +  \rho_1 \Phi_{\text{Summer}}^{(t-1)}  + \underbrace{\sum_{i=1}^3 \nu_i e^{t-i}}_{\text{new MA-term}} \\
        \tau^{(t)} & = \frac{\exp{(k\Phi_{\text{Winter}}^{(t)})}}{\exp{(k\Phi_{\text{Winter}}^{(t)})}+\exp{(k\Phi_{\text{Summer}}^{(t)})}}
    \end{split}
\end{equation}

\noindent \textbf{Model Parameters:}

\begin{equation}
    \begin{split}
        \text{UA}_0 & \sim \Gamma(4.75, 0.3) \\
        \text{UA}_{\text{W}} & \sim \Gamma(0.27,0.02) \\
        \text{T}_{\text{base}} & \sim \text{Normal}(18,1.2) \\
        \text{gA} & \sim \Gamma(4, 0.075)  \\
        \sigma_1 & = \Gamma(4.25, 0.82) \\
        \sigma_{\text{reduction}} & = \Gamma(4.25, 0.82) \\
        k & \sim \Gamma(1, 1)
    \end{split}
\end{equation}

\hrule
\end{minipage}
\end{table*}

%-----------------------------------------------------------
% Nomenclature TABLE - Updated
%------------------------------------------------------------	
\begin{table*}[h]
\small
\centering
\begin{tabularx}{\textwidth}{r|X|c}
\toprule
{\textbf{Name}} & {\textbf{Description}} & {\textbf{Unit}} \\
\midrule
\midrule
{} & {\textbf{Model Terminology}} & {} \\
\midrule
ES & Energy Signature model & \\
ARX-ES & ES model extended with an Auto-Regressive (AR) component & \\
ARMAX-ES & ARX-ES model extended with a Moving-Average (MA) component & \\
EPC & Energy Performance Certificate & \\

\midrule
{} & {\textbf{Heat Exchanger Data}} & {} \\
\midrule
Id & Heat exchanger tracking number & \\
$t$ & Date & yyyy-mm-dd \\
$\mathbf{\Phi}_t$ & Daily avg. heat consumption measurement & MW \\
$A$ & Total heated living area in building group & m$^2$ \\

\midrule
{} & {\textbf{Weather Data}} & {} \\
\midrule
$\mathbf{T}_{\text{ambient}}$ & Daily average ambient temperature & $^\circ$C \\
$\mathbf{W}_{\text{speed}}$ & Daily average wind speed & m/s \\
$\mathbf{I}_{\text{g}}$ & Daily average solar radiation per m$^2$ incident to earth's surface & W/m$^2$ \\
$t$ & Time and Date & yyyy-mm-dd hh \\

\midrule
{} & {\textbf{Model Parameters}} & {} \\
\midrule
HLC & Heat Loss Coefficient & kW/K \\
$\text{UA}_0$ & Base heat loss coefficient (wind-free conditions) & kW/K \\
$\text{UA}_{\text{wind}}$ & Wind infiltration coefficient & kW/(m$^2\cdot$K·(m/s)) \\
$T_{\text{base}}$ & Base temperature. & $^\circ$C \\
$gA$ & Solar gain factor & [–] \\
$k$ & Mixing parameter for regime transition (LSE function) & [–] \\
$\rho_i$ & $i^{\text{th}}$ Auto-Regressive (AR) coefficient & [–] \\
$\nu_i$ & $i^{\text{th}}$ Moving-Average (MA) coefficient & [–] \\

\midrule
{} & {\textbf{Model Output and Derived Quantities}} & {} \\
\midrule
$\Phi_{\text{winter}}$ & Weather-dependent heating regime output & MW \\
$\Phi_{\text{summer}}$ & Base-load (summer) regime output & MW \\
$\Phi_{\text{base}}$ & Base heat-load parameter & MW \\
$\tau(t)$ & Smooth weighting between noise parameters of seasonal regimes (LSE function) & [–] \\
$\mu(t)$ & Posterior predictive mean of heat demand & MW \\
$\sigma(t)$ & Posterior predictive standard deviation & MW \\
$\epsilon(t)$ & Error term for $\Phi_{\text{winter}}$ & MW \\
$\xi(t)$ & Error term for $\Phi_{\text{summer}}$ & MW \\

\bottomrule
\end{tabularx}
\caption{Nomenclature used throughout the article}
\label{tab:heat}
\end{table*}
%------------------------------------------------------------

\end{document}